\documentclass[pra,showpacs,preprintnumbers,amsmath,amssymb,tightenlines,twocolumn,superscriptaddress]{revtex4}

\usepackage{graphicx}
\usepackage{dcolumn}
\usepackage{bm}
\usepackage{epsfig}
\usepackage{stmaryrd}
\usepackage{color}

\newcommand{\bra}[1]{\langle\,{#1}\, |}
\newcommand{\ket}[1]{|\,{#1}\,\rangle}

\newcommand{\vek}[1]{\boldsymbol{#1}}
\newcommand{\eref}[1]{Eq.~(\ref{#1})}

\newcommand{\cref}[1]{chapter~\ref{#1}}

\newcommand{\Cref}[1]{Chapter~\ref{#1}}

\usepackage{ulem}  
\begin{document}

\title{Classical hallmarks of macroscopic quantum wave function propagation}

\author{James M.\ Feagin}
\affiliation{Department of Physics, California State University-Fullerton, Fullerton, CA 92834, USA }
\email{jfeagin@fullerton.edu}

\author{John S.\ Briggs}
\affiliation{Institute of Physics, University of Freiburg, Freiburg, Germany and \\
Department of Physics, Royal University of Phnom Penh, Cambodia}
\email{briggs@physik.uni-freiburg.de}

\begin{abstract}
The precise connection between quantum wave functions and the underlying classical trajectories often is presented rather vaguely by practitioners of quantum mechanics. Here we demonstrate, with simple examples, that the \emph{imaging theorem} (IT) based on the semiclassical propagator provides a precise connection. Wave functions are preserved out to macroscopic distances but the variables, position and momentum, of these functions describe classical trajectories. We show that the IT, based on an overtly time-dependent picture, provides a strategy alternative to standard scattering theory with which to compare experimental results to theory.
\end{abstract}

\maketitle

\section{Introduction}
In a previous paper \cite{BriggsFeagin_ITIII}  we showed that any system of particles emanating from microscopic separations describable by quantum dynamics will acquire characteristics of classical trajectories through normal unitary propagation to the macroscopic
separations at which measurements are made. This result we call the generalised \emph{imaging theorem} (IT). Here we will illustrate the operation of this formal IT result using simple examples demonstrating the direct imaging 
of quantum momentum distributions originating from a collision complex by measurement of the corresponding quantum spatial distribution of counts at a fixed remote detector as a function of time. We emphasise that this allows a method with which to compare directly experiment with theory, as alternative to the traditional specification of cross-sections. The IT provides an asymptotic spatial wave function, propagating according to the Schr\"odinger equation but whose coordinates develop according to classical Newton's equation. On the basis of the ensemble interpretation of the quantum wave function we show how these apparently contradictory features are reconciled.

We make clear that here we discuss only the unbounded motion of particles whose mass is such that, when the inter-particle distances are on the atomic scale, they must be treated by quantum mechanics. That is we confine discussion to quantum states in a continuum. Since all particles are composites, a single particle is defined as one for which the internal binding forces are stronger than external forces, so that the latter can be taken as acting on the centre-of-mass of the composite. Then we are discussing the motion of such particles from microscopic distances to the macroscopic separations at which detection is made. This is the typical situation in practically all collision processes in atomic, molecular, nuclear, and even high-energy physics. Particles and photons collide and one observes the many-particle fragmentation distant from the collision region. From our definition, internal degrees of freedom correspond to bound states of the composite particle. Hence they are quantized and must always be treated by quantum mechanics. 

The ideas that narrow wavepackets and Ehrenfest's theorem \cite{Ehrenfest} embody the nature of the quantum to classical transition pervade most text books on quantum mechanics. Conditions are sought in which a single material particle can be adequately represented by a packet of probability waves. This approach suffers from serious shortcomings, however. Firstly, the Ehrenfest theorem does not reproduce exactly Newton's equations of motion, except for certain simple potential forms. Secondly, the theorem involves the motion of the centre-of-mass of wavepackets which has little meaning for widely dispersed wavepackets. Thirdly, precisely this dispersion of wavepackets is unavoidable and leads to delocalisation except in the case of particles of macroscopic size. 
 
 The IT remedies these deficiencies in that the wave function and its inevitable dispersion are preserved under propagation to large distances and times. Furthermore, the IT depends upon a transition of the quantum propagator to semiclassical form, which is valid for values of the action greater than $\hbar$ for any potential function. In the IT it is the {\it{variables}} of quantum wave functions which obey classical mechanics in their time variation, not simply the centre-of-mass of a wavepacket.

The IT provides also a concrete mathematical justification for the common practice used by experimentalists to assume classical mechanics, even for electrons, to describe the motion of particles from a microscopic reaction zone to a detector at macroscopic distances. The classical trajectory, often guided by external fields, is used to propagate backwards in time to infer initial momentum at the edge of the reaction zone from ``hits" measured at a given position on the detector at a given time relative to the instant of reaction. In this way a measured ``time spectrum" is compared to the cross section differential in particle momenta, as provided by standard \emph{time-independent} scattering theory.
 
 Increasingly over the last years, following the development of femtosecond (molecular) and recently attosecond (atomic) timing techniques, attention is turning to studying the time development of collision processes. This is true particularly in the field of fragmentation by short laser pulses \cite{Maquet}, demanding a comparison with time-dependent scattering theory. The IT is well suited to the task. As will be shown below, it provides a direct comparison of calculated quantum probabilities propagated in time with measured time spectra of detector counts at fixed macroscopic positions. This is the essence of the imaging of quantum wave functions at the edge of the microscopic reaction zone.
 
 In section II we begin with a brief exposition of the main results of Ref.\ \cite{BriggsFeagin_ITIII}. Then we show how the probabilities calculated from time-dependent positional wave functions can be compared directly to time spectra measured at fixed positions on a remote detector. The key ingredient of the theory is that the time development of quantum position variables outside the reaction zone follows classical mechanics. (We assume however the motion is nonrelativistic.) The locus of equal probabilities is a classical trajectory. The various classical actions with which to construct the semiclassical wave function are then described for the cases of free propagation and motion in a constant extracting field. Despite these classical aspects, we demonstrate for the simple example of photofragmentation of 
 the $H^+_2$ molecular ion, that all aspects of the quantum wave function, in particular the nodal structure, are preserved in the propagation to a detector at macroscopic distance.

The main results are summarised in the Conclusions section where we comment upon the apparent dichotomy posed by the IT. A quantum wavefunction gives rise to a probability distribution corresponding to an ensemble of particles moving on classical trajectories.

 \section{The imaging theorem }
The IT in the one-body asymptotically free case is extremely simple. Consider a particle emitted from a reaction zone at time $t = t_i$ 
near a point $\vek r(t_i) = \vek r_i$ close to the origin
with a momentum distribution described by a 
 momentum-space wave function $\tilde\Psi(\vek p, t_i)$. The particle propagates a macroscopic distance to a detector at $\vek r(t_f) = \vek r_f$, with $r_f \gg r_i$. The position-space wave function at the detector is given by
 \begin{equation}
 \label{exactwfn1}
\Psi(\vek r_f,t_f)  = \int  d\vek p \, \tilde K(\vek r_f,t_f; \vek p,t_i) \,  \tilde\Psi(\vek p, t_i),
\end{equation}
where $\tilde K(\vek r_f,t_f; \vek p,t_i) = \bra{\vek r_f}U(t_f,t_i)\ket{\vek p}$ is the mixed coordinate-momentum propagator and  $U(t,t_i)$ is the time-development operator.  For free motion 
of a particle of mass $m$ the propagator has already the semiclassical form, and one has specifically the Fourier transform integral
 \begin{equation}
 \label{exactprop}
\Psi(\vek r_f,t_f)  = (2\pi\hslash)^{-3/2} \int  d\vek p \, e^{i \vek p \cdot \vek r_f /\hslash- i p^2 t/(2m\hslash)} \,\tilde\Psi(\vek p, t_i),
\end{equation}
where $t \equiv t_f - t_i$.
 Important to note here is the wave property that the amplitude of the wave function at position $\vek r_f$ depends upon {\it{all}} values of the initial momentum $\vek p$.

The integral is evaluated in stationary phase approximation. The stationary phase point occurs then at the free motion classical value $\vek p_i = m \vek r_f/t$ and one derives, for asymptotically large $\vek r_f$ and $t$ but such that  the velocity $r_f/t $ remains constant,
 \begin{equation}
 \label{ITwavefree}
\Psi(\vek r_f, t_f)  \approx \left(\frac{m}{i t}\right)^{3/2} e^{ i m r_f^2/(2\hbar t)} \, \tilde\Psi(\vek p_i, t_i).
\end{equation}
This is the IT for single particle free asymptotic motion. Important is that the amplitude of the wave function  at each final $\vek r_f$ is connected by a {\it{classical trajectory}} to the unique fixed initial momentum $\vek p_i$. There is no longer an integral over all $\vek p$ values as in \eref{exactprop}. The transition to classical mechanics occurs in the arguments of the wave functions. For example, the familiar spreading of the wave function in time is linked to the natural separation in time of classical trajectories with different initial momenta $\vek p_i$. The wave functions themselves are preserved giving rise to possible quantum effects. This mixed quantum--classical character is the hallmark of the IT.
 
 If one notes that $(m/t)^{3/2} = (d\vek p_i/d\vek r_f)^{1/2}$, the IT can be written in the form
  \begin{equation}
  \label{locus}
|\Psi(\vek r_f,  t_f)|^2\,d\vek r_f =  |\tilde\Psi(\vek p_i, t_i)|^2\, d\vek p_i.
\end{equation}
This remarkable result emphasises the ensemble picture of quantum mechanics. A distribution of particles with different momenta $\vek p_i$ emanates from a reaction zone and propagates in such a way that the locus of points of equal detection probability follow classical trajectories. Although the IT of \eref{ITwavefree} was derived originally by Kemble in 1935 \cite{Kemble} and the result is more important than Ehrenfest's theorem \cite{Ehrenfest}, sadly it has not found its way into quantum text books.

In Ref.\ \cite{BriggsFeagin_ITIII} we generalised the IT to describe any number of particles, possibly interacting between themselves and being extracted by external classical fields. 
The generalisation of \eref{ITwavefree} to $n$ particles is 
\begin{eqnarray}
 \label{Psixftf}
 \Psi(\vek r_f, t_f) &\approx& \,(-i)^{3n/2}\,\left( \frac{ d\vek p_i}{d\vek r_f} \right)^{1/2} \nonumber  \\ 
 	&\times& \exp\left(\frac{i}{\hslash} S_c(\vek r_f,t_f; \vek r_i, t_i)  \right)  \tilde\Psi(\vek p_i, t_i)
\end{eqnarray}
where $\vek r_f$, $\vek r_i$, and $\vek p_i$ are $n$-dimensional position and momentum vectors which include all the particles and $S_c$ is the classical action function. 
 
 Again one can express this general IT in the form
 \begin{equation}
\label{VVtheorem2}
|\Psi(\vek r_f, t_f)|^2 \approx \frac{d \vek p_i}{d \vek r_f} \, |\tilde\Psi(\vek p_i, t_i)|^2,
\end{equation}
 which is also \eref{locus} for $n$ particles. This form has a wholly classical, statistical interpretation. An ensemble of particles with probability density $ |\tilde\Psi(\vek p_i, t_i)|^2$
 of initial momentum $\vek p_i$ move on classical trajectories and hence are imaged at later times as the position probability density $|\Psi(\vek r_f, t_f)|^2$. 
 The factor $d \vek p_i/d \vek r_f$ is the classical trajectory density of finding the system in the volume element $d\vek r_f$ given that it started with a momentum $\vek p_i $ in the volume element $d \vek p_i$ (see Gutzwiller \cite{semicl}, chap.\ 1). Quantum mechanics merely furnishes the initial momentum distribution. 

We note that the IT is based on the limit $r_f \gg r_i$ along with $t_f \gg t_i$. Particles emanating from a microscopic reaction zone  are detected at macroscopic distance. Nevertheless, one can formally and arbitrarily include a small $\vek r_i$ with the replacement of the momentum wave function on the RHS of \eref{exactwfn1} by $e^{-i \vek p \cdot \vek r_i} \tilde \psi(\vek p, t_i)$ to shift the spatial origin of the state to $\vek r_i$ to recognize an explicit starting point for the classical trajectory near to the boundary of the reaction volume. Practically, since detection is at macroscopic distances and all particles emanate from a volume of atomic dimensions, one can take $\vek {r}_i = 0$ without loss of accuracy.

 In the following section we will show that the IT implies that all information on the scattering process is contained in the detection of  the number of particles arriving at a remote detector within a volume element $d\vek{r}_f$ as a function of the time of flight $t = t_f - t_i$. As we shall demonstrate, this is precisely the \emph{time spectrum} measured by experiment.
 This number is proportional to the spatial distribution $|\Psi(\vek r_f, t_f)|^2$ and the IT connects this to the initial momentum distribution $|\tilde \Psi(\vek p_i,t_i)|^2$, which is equal to the modulus squared of the momentum-space $T$ matrix element of scattering theory. 
Then we consider the form of the classical actions corresponding to free motion and motion in a uniform electric or gravitational field in order to construct $|\Psi(\vek r_f, t_f)|^2$  from a given $|\tilde \Psi(\vek p_i,t_i)|^2$. Finally we illustrate this procedure in detail with a model of photodissociation of the $H_2^+$ molecular ion and detection of both $H$ and $H^+$ fragments.

\subsection{Position Detection}

The time spectrum is the primary measured element of most modern scattering experiments 
\cite{energyspectra}. Substituting $t \rightarrow t(\vek r_f, \vek p_i)$ from the classical trajectory as a function of the final detected position and the initial momentum and dividing by the classical density $d\vek p_i/d\vek r_f$, we obtain from \eref{VVtheorem2} for fixed $\vek r_f$ 
\begin{equation}
\label{phiITsq}
|\tilde \Psi(\vek p_i,t_i)|^2 \approx \left(\frac{d\vek p_i}{d\vek r_f} \right)^{-1} |\Psi(\vek r_f, t_f)|^2   \, 
	\rule[-2mm]{.1mm}{6mm}_{\,t \rightarrow t(\vek r_f, \,  \vek p_i)}.
\end{equation}
This IT approximation becomes exact for $\vek r_f$ and $t$ large enough and certainly for the macroscopic parameters of a typical laboratory apparatus. 
We will examine examples of this result in the following sections.

Traditionally in collision physics the theorist calculates a cross section (differential or total) in terms of final momenta and the experimentalist converts the measured data (expressed in terms of the flux of particles) to appropriately confront experiment with theory. Modern multi-particle coincident detectors measure directly time spectra, not momentum or energy, of the number of particles detected at a particular position over a given collection time. The initial ejection momentum is then ascertained by {\it{assuming}} classical particle motion from the microscopic reaction zone to the detector at macroscopic distance away. The IT is the justification of this step. As measurements become more sophisticated in the number of particles measured and the degree to which the measurement is differential in the momentum coordinates, the way in which to compare experiment with theory becomes increasingly complicated. In a sense the IT offers the alternative in allowing the counts per unit time into a small volume $d\vek r_f$ on the detector to be calculated directly from theory. The method is similar to that proposed already by Macek and co-workers as a method to extract data from numerically-propagated many-particle time-dependent wave functions \cite{Macek}.

The essence of the IT for scattering theory is that the momentum wave function $\tilde\Psi(\vek p_i, t_i)$ in \eref{Psixftf} is identical to the usual end product of a scattering theory, the $T$-matrix element in momentum space. Hence we put $\tilde\Psi(\vek p_i, t_i) \equiv T(\vek p_i)$. Then we use the IT of \eref{phiITsq} to relate the detected time spectrum $ |\Psi(\vek r_f, t_f)|^2$ directly to the modulus square of the $T$ matrix. 
This strategy circumvents the definition of a  multi-dimensional differential cross-section for a particular process. However, in the usual way, if some particles or some dynamical variables (e.g.\ precise direction)
of a given particle are not detected then an appropriate integral over these variables must be made.

We note in passing that the semiclassical wave function \eref{ITwavefree} is an eigenfunction of the quantum momentum operator but with an eigenvalue given by the classical particle momentum at the detector $\vek p(t_f) = \vek p_f$ \cite{BriggsFeagin_ITII}.
It follows that the quantum probability current density at the detector in the IT limit is proportional to the classical velocity at the target $\vek v_f =  \vek p_f/\mu$, viz.\
\begin{equation}
\vek j = \mbox{Re}\{\Psi^*(\vek r_f,t_f) \frac{1}{\mu} \vek p \, \Psi(\vek r_f,t_f) \} \sim  |\Psi(\vek r_f,t_f)|^2 \, \vek v_f. 
\label{jdensity}
\end{equation}
This result is just another statement of the IT.

 \subsection{The classical actions}
 To keep the development simple, we will consider  the case of individual particle motion in only one-dimension, both free and in the presence of a constant force in the asymptotic region. This corresponds to the common electric field extraction of charged particles or to the free fall in the gravitational field. In the case of free motion, the classical action occurring in 
 \eref{Psixftf}, now denoted by $S_0$, for propagation from initial position $z_i$ to final position $z_f$ is
 \begin{equation}
 \label{S0zi}
S_0(z_f,t_f; z_i,t_i) = m(z_f - z_i)^2/(2t)
\end{equation}
where $t = t_f - t_i$. However, the kernel in \eref{exactwfn1} is proportional to the mixed coordinate-momentum action $\tilde S_0(z_f,t_f; p,t_i)$ where we denote the initial momentum to be integrated over as $p$.  
 This action is obtained by the Legendre transformation
\begin{equation}
\label{S0pi}
\tilde S_0(z_f,t_f; p,t_i) = S_0(z_f,t_f; z_i,t_i) + pz_i = pz_f - p^2t/(2m).
\end{equation}
 This is the action appearing in \eref{exactprop}.

  The generalisation to motion in a constant force $F$ is straightforward. The coordinate action $S_F(z_f,t_f; z_i,t_i)$ is \cite{BriggsFeagin_ITII}
\begin{equation}
\label{K_F}
S_F(z_f,t_f; z_i, t_i) = F t z_f - \frac{F^2 t^3}{6m} + \frac{m}{2t} \left[z_f - z_i -  \frac{F t^2}{2m} \right]^2,
\end{equation} 
which reduces to \eref{S0zi} in the $F=0$ limit.
Performing a Legendre transformation as in \eref{S0pi} with the initial position from the classical trajectory,
$ z_i = z_f - p t/m - F t^2/(2m),$
one obtains
\begin{equation}
\label{SFp1}
\tilde S_F(z_f,t_f; p, t_i) = (p + F t)\left(z_f - \frac{F t^2}{2m} \right) - \frac{F^2 t^3}{3m} - \frac{p^2 t}{2m},
\end{equation}
which reduces to \eref{S0pi} in the $F = 0$ limit.
For $F$ finite, one defines the stationary phase with
\begin{equation}
\label{SF_SP}
\frac{\partial \tilde S_F}{\partial p}  = z_f - \frac{p t}{m} - \frac{F t^2}{2m} = z_i \equiv 0.
\end{equation}
This is just the condition appropriate for the IT limit $z_f \gg z_i$.
Thus one obtains as the point of stationary phase the initial momentum from the classical trajectory,
\begin{equation}
\label{pi_SP}
p \rightarrow \frac{m}{t} \left(z_f - \frac{F t^2}{2m} \right) \equiv p_i.
\end{equation}

\section{Detection of $H_2^+$ fragmentation}

Continuing with one dimension for simplicity, we consider a specific experiment in which a beam of $H_2^+$ molecules in the ground vibrational state of the ground electronic state is crossed with a laser of sufficient energy to dissociate the molecule into  $H + H^+$. 
Then one can detect $H$ atoms moving freely asymptotically, or use a constant electric field to divert $H^+$ ions onto a detector a macroscopic distance away.
We shall also show how the neutral $H$ atom detection can be enhanced, analogous to electric field extraction, by a momentum boost of the center of mass (CM) of the $H + H^+$ pair.
 
The analysis is made conveniently by considering harmonic oscillator states as a good approximation to the $H_2^+$ vibrational states. Then the free propagation wave functions can be calculated exactly. 
The initial vibrational states describing the $H + H^+$ relative motion in harmonic approximation have the momentum wave functions

\begin{equation}
\label{phi_nt}
\tilde\Psi_n(p) = \frac{i^{-n}}{\sqrt{2^n n!}} \frac{e^{-p^2/(2 \mu \hslash \omega)}}{(\pi \mu \hslash \omega)^{1/4}}  H_n(\frac{p}{\sqrt{ \mu \hslash \omega}})
\end{equation}
defined by the reduced mass $\mu = m_p/2$ ($m_p$ is the proton mass) and an effective vibrational frequency $\omega$ with $H_n(z)$ a Hermite polynomial.
With this initial state,  \eref{exactprop} (in 1D) can be evaluated exactly by completing the square on $p$ in the exponent 
and invoking a standard integral  \cite{GradRyzh}.
One obtains for the free propagation of \eref{phi_nt}
\begin{equation}
\label{psi_nt}
\begin{split}
\Psi_n(z_f, t_f) &= \frac{i^{-n}}{\sqrt{2^n n!}} \left( \frac{\mu \omega}{\pi \hslash} \right)^{1/4}  \left (\frac{-1 + i \omega t}{ 1 + i \omega t} \right)^{n/2}  \\ 
	&\times \frac{e^{-\mu \omega z_f^2/(2 \hslash (1 + i \omega t))}}{\sqrt{1 + i \omega t}}  H_n \left(\sqrt{\frac{\mu \omega}{\hslash (1 + \omega^2 t^2)}} \, z_f \right)
\end{split}
\end{equation}
for  $t = t_f - t_i$. 
For large times such that $\omega t \gg 1$, this exact result becomes
\begin{equation}
\label{psiIT_nt}
\begin{split}
\Psi_n(z_f, t_f)  &\approx \left(\frac{\mu}{ i t}\right)^{1/2} e^{i \mu z_f^2/(2\hslash t)}  \\ 
&\times  \frac{i^{-n}}{\sqrt{2^n n!}} \frac{e^{-(\mu z_f/t)^2/(2 \mu \hslash \omega)}}{(\pi \mu \hslash \omega)^{1/4}} H_n \left(\frac{\mu z_f/t}{\sqrt{ \mu \hslash \omega}} \right)\\
&=  \left(\frac{\mu}{ i t}\right)^{1/2} e^{ i \mu z_f^2/(2 \hslash t)} \, \tilde \Psi_n(\mu z_f/t),
\end{split}
\end{equation}
which is the precise 1D form of the IT of  \eref{ITwavefree}. 
Of course the same result is obtained by evaluating the Fourier transform integral in \eref{exactprop}  (in 1D) in stationary phase approximation around the stationary phase and classical trajectory point $p \rightarrow p_i = \mu z_f /t$. 

The three-dimensional harmonic-oscillator wavepackets are simply products for the $(x,y,z)$ directions and the IT gives the general result \eref{VVtheorem2} which shows that asymptotically, for detection of single particles 
described by harmonic-oscillator wavepackets, there is no difference between quantum and classical ensembles. The result is true for particles of arbitrary mass, there is no need to go to the limit of particles of macroscopic mass and wavepacket widths less than the size of the particle. In this sense, the IT re-emphasises that quantum mechanics is only concerned with statistical ensembles and cannot describe single particles.
\begin{figure}[t]
\includegraphics[scale=.255]{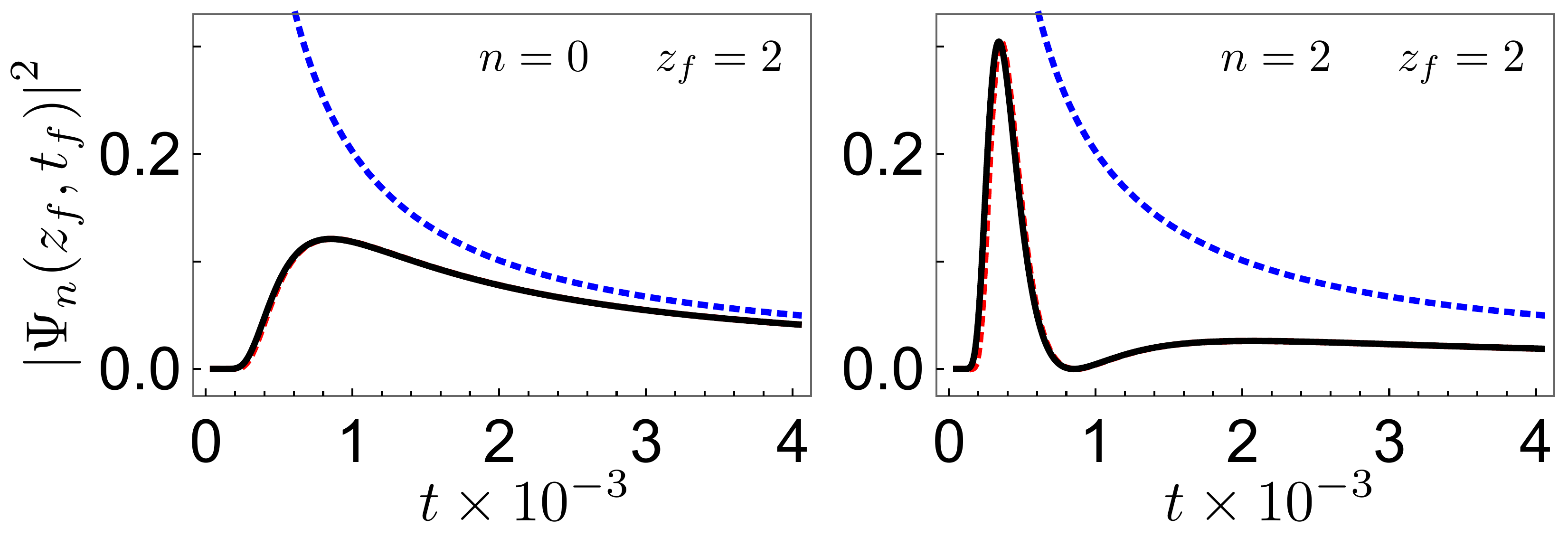}
\caption{\label{fig1} Time spectra from the spatial distributions $|\Psi_{n}(z_f, t_f)|^2$ for fixed $z_f$ as a function of $t = t_f-t_i$. The solid black curves show the exact distributions from \eref{psi_nt} while the dotted red curves show the IT limit from \eref{psiIT_nt}.
The dotted blue curves show the classical density $\mu/t$.}
\end{figure}

The spatial distribution $|\Psi_{n}(z_f, t_f)|^2$ for fixed $z_f$ as a function of $t = t_f-t_i$ defines a time spectrum of detector hits. Two such spectra for $n = 0$ and $2$ are illustrated in Fig.\ \ref{fig1} in atomic units (au) with $\hslash = 1$, $\mu = 918$,   and $\omega = 0.01$ estimated from a formula for the $H_2^+$  vibrational spectrum \cite{Pauling-Wilson}. ($1 \, \mbox{au of time} \approx 2.42 \times 10^{-17} \, \mbox{s}$.)

To demonstrate the rapid convergence of the IT to the exact results we place the detector at $z_f = 2 \, a_0$ ($a_0 \, \mbox{ the Bohr radius} \equiv 1  \, \mbox{au of length} \approx 5.29 \times 10^{-11} \, \mbox{m}$), a microscopically small distance from the origin but nevertheless somewhat beyond the range of the initial spatial distributions $|\Psi_{n}(z_f, 0)|^2$. 
For small $t$, the time spectrum is vanishing because the wave function propagating out from the reaction zone has not yet reached the detector.
Clearly the classical density $dp_i/dz_f = \mu/t$ defines the overall trend of the time spectrum and approaches asymptotically the quantum density. 

\begin{figure}[b]
\includegraphics[scale=.265]{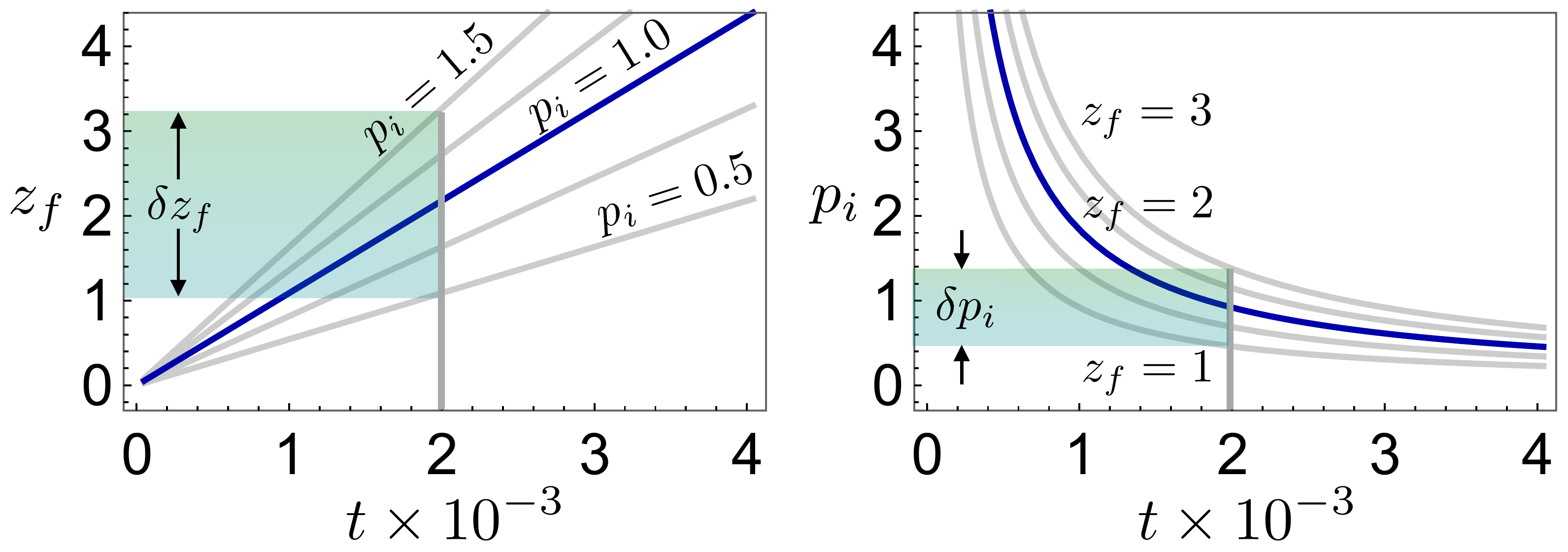}
\caption{\label{fig1p} Classical free trajectories $z_f$ (left panel) and $p_i$ (right panel) plotted as a function of time about the nominal values (blue curves) $p_i = 1$ and $z_f = 2$, repectively. The green rectangles indicate ranges $\delta z_f$ and $\delta p_i$ at $t = 2000$.}
\end{figure}

In Fig.\ \ref{fig1p}, we illustrate the origin of the classical density for free motion in 1D. We show corresponding fans of classical trajectories for $z_f = p_i t/\mu$ and $p_i = \mu z_f/t$ in atomic units. One readily sees that the range $\delta z_f$ increases as $t$ increases in proportion to a decrease in the range $\delta p_i$, as specified by the classical density $\delta p_i/\delta z_f \sim \mu/t$.

In practice, experimentalists convert the time spectrum to an energy spectrum assuming classical kinematics and integrate it in sectors to define cross sections of various dynamical features. The IT connects the time spectrum to the initial momentum distribution $|\tilde \Psi_n(p_i)|^2$ directly. 
Substituting $t \rightarrow \mu z_f/p_i$ from the classical trajectory, we obtain from \eref{psiIT_nt} for fixed $z_f$ the 1D analog of \eref{phiITsq},
\begin{equation}
\label{phiIT1Dsq}
|\tilde \Psi_n(p_i)|^2 \approx \frac{t}{\mu} |\Psi_n(z_f, t_f)|^2   \, 
	\rule[-2mm]{.1mm}{6mm}_{\,t \rightarrow \mu z_f/p_i},
\end{equation}
which, essentially on a microscopic scale but outside the reaction volume, becomes effectively exact as $z_f$ and $t$ are increased. 
\begin{figure}[t]
\includegraphics[scale=.256]{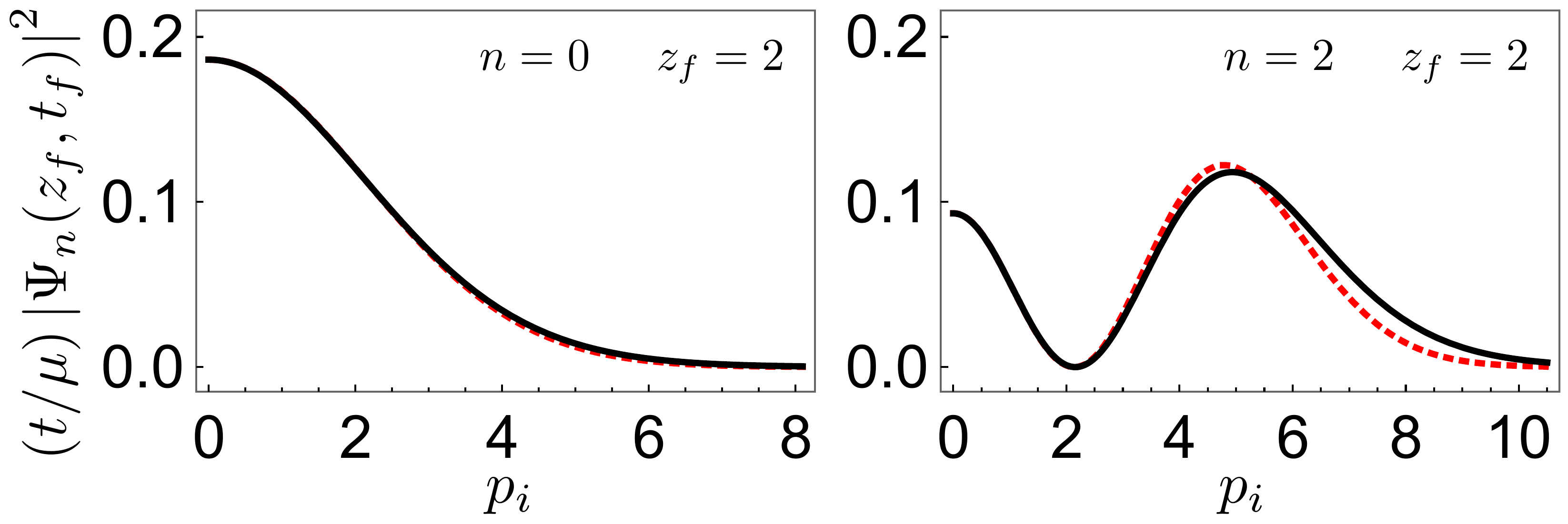}
\caption{\label{fig2} Initial momentum distributions extracted from the time spectra in Fig.\ \ref{fig1}. The solid black curves show the exact distributions from \eref{phi_nt} while the dotted red curves show the extracted results using \eref{phiIT1Dsq}.}
\end{figure}

We demonstrate convergence of this result in Fig.\ \ref{fig2} for the time spectra shown in Fig.\ \ref{fig1}. 
These extracted momentum distributions show $p_i > 0$ only.
With a detector placed along the $+z$ axis, one detects only particles with positive initial momentum. Particles emitted along the $-z$ axis go undetected. We demonstrate in the following sections how this can be remedied kinematically or with external-field extraction. 
The poorer convergence of the IT limit to the exact result in Fig.\ \ref{fig2} for large momentum is related to the poorer convergence in Fig.\ \ref{fig1} for small $t$ of the IT limit in the time spectrum: the fastest particles and hence the largest momenta are detected first. In any case, the IT approximation quickly improves with increasing $z_f$.

In the 1D examples we consider here, the probability current density \eref{jdensity} is simply the rate of detector hits. Our initial momentum states \eref{phi_nt} have definite parity and therefore their momentum densities are symmetric about the origin so that 
 $\int_{0}^{\infty}  j_n \, dt = 1/2$, since half the particles are emitted along the $-z$ axis and go undetected.
In fact one can show using \eref{psi_nt} and with $v_f = z_f/t$ that the ratio
\begin{equation}
\frac{j_n(z_f,t_f)}{|\Psi_n(z_f,t_f)|^2 \,  v_f} = \frac{\omega^2 t^2}{1+\omega^2 t^2}
\label{equ:jndensity}
\end{equation} 
independently of the initial state $n$. As required by the IT, the ratio approaches unity for $\omega t \gg 1$.

\subsection{Detection enhancement via a CM momentum boost}

To enhance detection of $p_i < 0$ particles and also to make contact with the text book picture of single-particle propagation we consider that the experiment by design imparts a momentum boost $p_c$ to the $H + H^+$ CM along the $+z$ axis. 
If one boosts the CM motion enough, one can collect even particles released with $p_i < 0$. This is easy to understand if one imagines sitting at the CM and watching the detector coming towards you with momentum $-p_c$. If the detector is moving fast enough it will always catch up with all the reaction fragments going away from you, even those that depart along the $-z$ axis.
The technique works for both neutral and charged particle extraction and has actually been implemented and refined by Helm and coworkers over the past decade to study laser dissociation of $H_3 \rightarrow 3H$ \cite{Helm2014}. 

If the $H$ atom is released with a momentum $p_i$ relative to the $H^+$, then it will strike a laboratory detector located at $z_f > 0$ with momentum $p = p_i + p_c/2 > 0$ if $p_c$ is large enough. Just how large is determined by the range of the initial momentum distribution $ |\tilde\Psi(p_i,t_i)|^2$ we are looking to extract. 
Therefore we introduce $ \tilde\Psi(p-p_c/2, t_i)$ in \eref{exactprop} (in 1D) to derive the formal time development of the boosted spatial wave function in the laboratory frame, denoted by a subscript $c$,
\begin{equation}
\label{psiboost}
\Psi_c(z_f, t_f) = e^{i p_0 z_f/(2\hslash) - i p_0^2 t/(2\mu \hslash)} \Psi(z_f - p_0 t/\mu, t_f)
\end{equation}
with $p_0 \equiv p_c/2$ and $t = t_f - t_i$.
In the IT limit $\omega t \gg 1$ we obtain with \eref{psiIT_nt} and the replacement $z_f \rightarrow z_f - p_0 t/\mu$,
\begin{eqnarray}
\label{psiITpc}
\Psi_{nc}(z_f, t_f) &\approx& e^{i p_0 z_f/\hslash - i p_0^2 t/(2\mu \hslash)} \nonumber \\
	&\times&  \left(\frac{\mu}{ i t}\right)^{1/2} e^{ i \mu (z_f - p_0 t/\mu)^2/(2 \hslash t)} \tilde \Psi_n(p_i), \nonumber \\
	&=&  \left(\frac{\mu}{ i t}\right)^{1/2} e^{ i \mu z_f^2/(2 \hslash t)} \tilde \Psi_n(p_i),
\end{eqnarray}
with $p_i \equiv \mu z_f /t - p_c/2$.
The same result is obtained from \eref{exactprop} (in 1D) inserting $ \tilde\Psi_n(p-p_c/2)$ and evaluating the integral in stationary phase approximation around the same free-motion stationary phase point $p \rightarrow \mu z_f/t$. 

\begin{figure}[t]
\includegraphics[scale=.256]{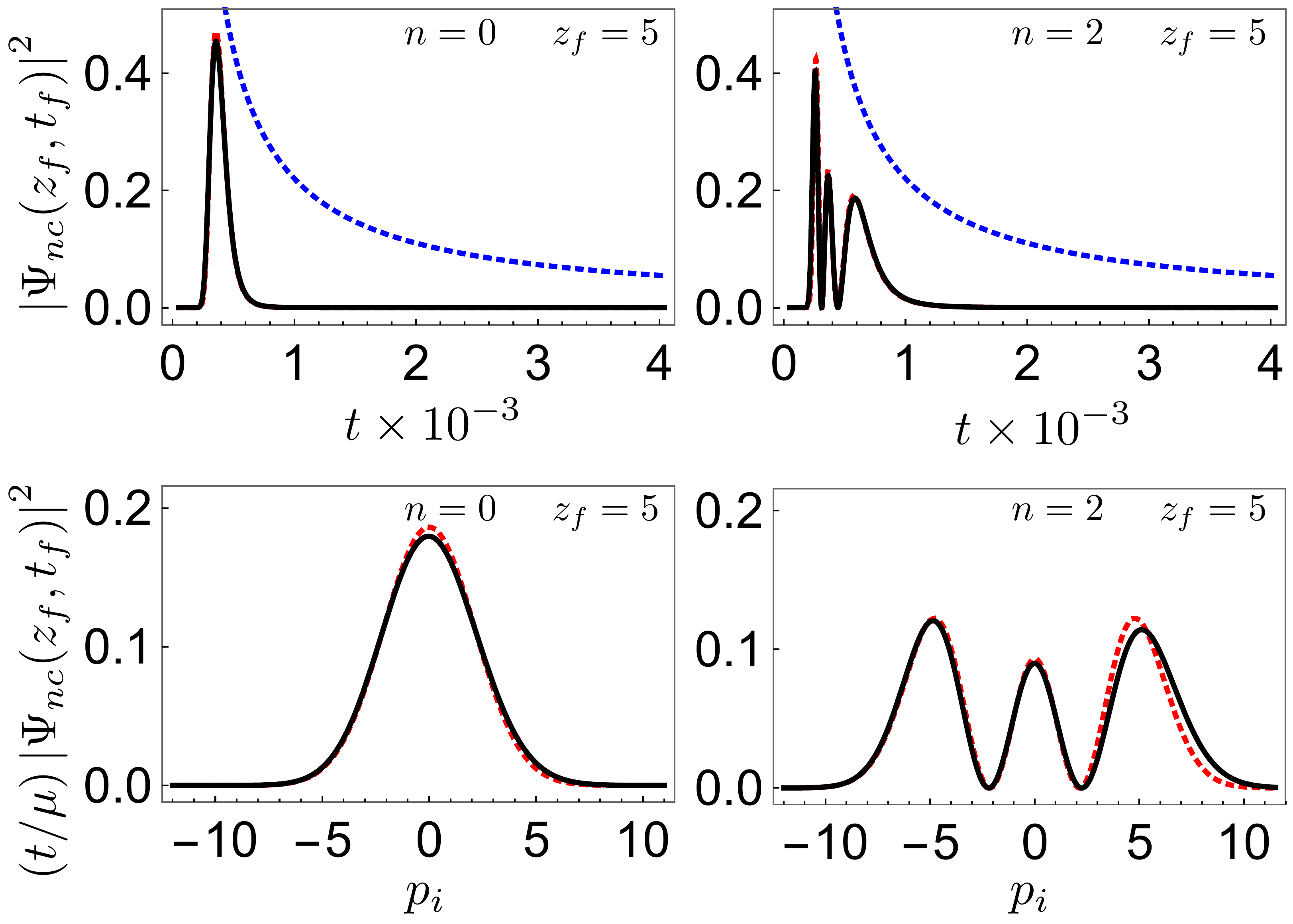}
\caption{\label{fig3} Top two panels show time spectra from the spatial distributions $|\Psi_{nc}(z_f, t_f)|^2$ for fixed $z_f$ as a function of $t = t_f-t_i$. The solid black curves show the exact distributions from Eq.\ (\ref{psi_nt}) using Eq.\ (\ref{psiboost}) while the dotted red curves show the IT limit from \eref{psiITpc}. The dotted blue curves show the classical density $\mu/t$. Bottom two panels show initial momentum distributions extracted from the time spectra. The solid black curves show the exact distributions from \eref{phi_nt} while the dotted red curves show the extracted results using \eref{phiITpcsq}.}
\end{figure}

Fig.\ \ref{fig3} shows the time spectra of Fig.\ \ref{fig1} boosted by the CM momentum $p_c =  25 \, v_0$ using Eq.\ (\ref{psiboost}). ($v_0 \, \mbox{ the Bohr velocity} \equiv 1  \, \mbox{au of velocity} \approx 2.19 \times 10^6 \, \mbox{m/s}$.) Here the detector has been moved out to $z_f = 5 \, a_0$ to show the $n=2$ nodes better.
With the momentum boost, the time spectra now resemble more the actual spatial harmonic distributions. For example, the $n=2$ spectrum exhibits the two nodes seen in the corresponding spatial distribution. 
Again, we see that the overall magnitude of the time spectrum is well described by the classical density.

In 1D, a single detector positioned at fixed $z_f$ can be used to extract via the IT the initial momentum distribution $|\tilde \Psi_n(p_i)|^2$ but including particles released along the $-z$ axis with $p_i < 0$. With $t \rightarrow \mu z_f/(p_i + p_c/2)$ from the classical trajectory, we obtain from \eref{phiITsq} for fixed $z_f$ 
\begin{equation}
\label{phiITpcsq}
|\tilde \Psi_n(p_i)|^2 \approx \frac{t}{\mu} |\Psi_{nc}(z_f, t_f)|^2   \, 
	\rule[-2mm]{.1mm}{6mm}_{\,t \rightarrow \mu z_f/(p_i + p_c/2)}.
\end{equation}
We demonstrate the enhanced detection and convergence of this result in Fig.\ \ref{fig3}.
Most notable, the full initial momentum distribution has been extracted for all $p_i$ and the integrated current density from \eref{phi_nt} now gives $\int_{0}^{\infty}  j_{nc} \, dt = 1$. Otherwise, convergence details are essentially the same as those described in connection with Figs.\ \ref{fig1} and \ref{fig2}.

Again, one can show using Eqs.\ (\ref{psi_nt}) and (\ref{psiboost}) and with $v_f = z_f/t$ that
\begin{equation}
\frac{j_{nc}(z_f,t_f)}{|\Psi_{nc}(z_f,t_f)|^2 \,  v_f} = \frac{p_0/p_f + \omega^2 t^2}{1+\omega^2 t^2}
\label{equ:jndensity}
\end{equation} 
($p_0 \equiv p_c/2$ and $p_f \equiv \mu v_f$) independently of the initial state $n$ as before. 
And as required by the IT, this ratio approaches unity for $\omega t \gg 1$.

\subsection{Detection enhancement via electric-field extraction}
   
One can establish a comparable extraction enhancement by introducing a constant force, for example a constant electric field to steer the ejected $H^+$, or a gravitational field in the case of gravity interferometry \cite{Greenberger}.  This constant-force action $S_F$ in \eref{K_F} is essentially a coordinate-translated version of the free-particle action $S_0$. Hence the accelerated state evolves as a Galilean-like boost of the free propagation description analogous to \eref{psiboost}  and takes on the exact form \cite{BriggsFeagin_ITII}
\begin{equation}
\label{PsiF}
\Psi_F(z_f,t_f) =  e^{i F t \, z_f/\hslash - i F^2 t^3/(6\mu \hslash)} \, \Psi(z_f-Ft^2/(2\mu),t_f).
\end{equation}  
In the IT limit $\omega t \gg 1$ we obtain with \eref{psiIT_nt} and the replacement $z_f \rightarrow z_f-Ft^2/(2\mu)$,
\begin{equation}
\label{PsiFIT}
\begin{split}
\Psi_{nF}(z_f,t_f) & \approx e^{i F t \, z_f/(2\hslash) - i F^2 t^3/(24\mu \hslash)}  \\
&\times  \left(\frac{\mu}{i t}\right)^{1/2} \exp\left[i \frac{\mu z_f^2}{2\hslash t}\right] \tilde \Psi_n(p_i),
 \end{split}
\end{equation}  
where now $p_i = \mu[z_f -  F t^2/(2\mu)]/t$, which is also
the stationary phase point from \eref{pi_SP}. 

Solving \eref{pi_SP} for $t$ gives
\begin{equation}
\label{tF}
t =  -\frac{p_i}{F} + \frac{\sqrt{p_i^2 + 2\mu F z_f}}{F},
\end{equation}  
which can be used in \eref{phiITsq}  to extract the initial momentum distributions from the time spectra as in the previous section.

As a demonstration of this procedure and a further test of the IT, we simulate a real experiment to extract the $H^+$ ions with an electric field. We use rejection sampling \cite{Koonin} to generate a spectrum of some $ 10^4$ random time of flight values $t = t_f-t_i$ distributed according to $|\Psi_F(z_f,t_f)|^2$ from \eref{PsiF} assuming an $n=2$ initial vibrational state from Eq.\ (\ref{psi_nt}). We use the same parameters as in the previous sections except here we place the detector at the macroscopic distance $z_f = 20 \, \mbox{cm}$. The resulting list of time values represents actual random detector clicks over a macroscopic time interval. 
\begin{figure}[b]
\includegraphics[scale=.255]{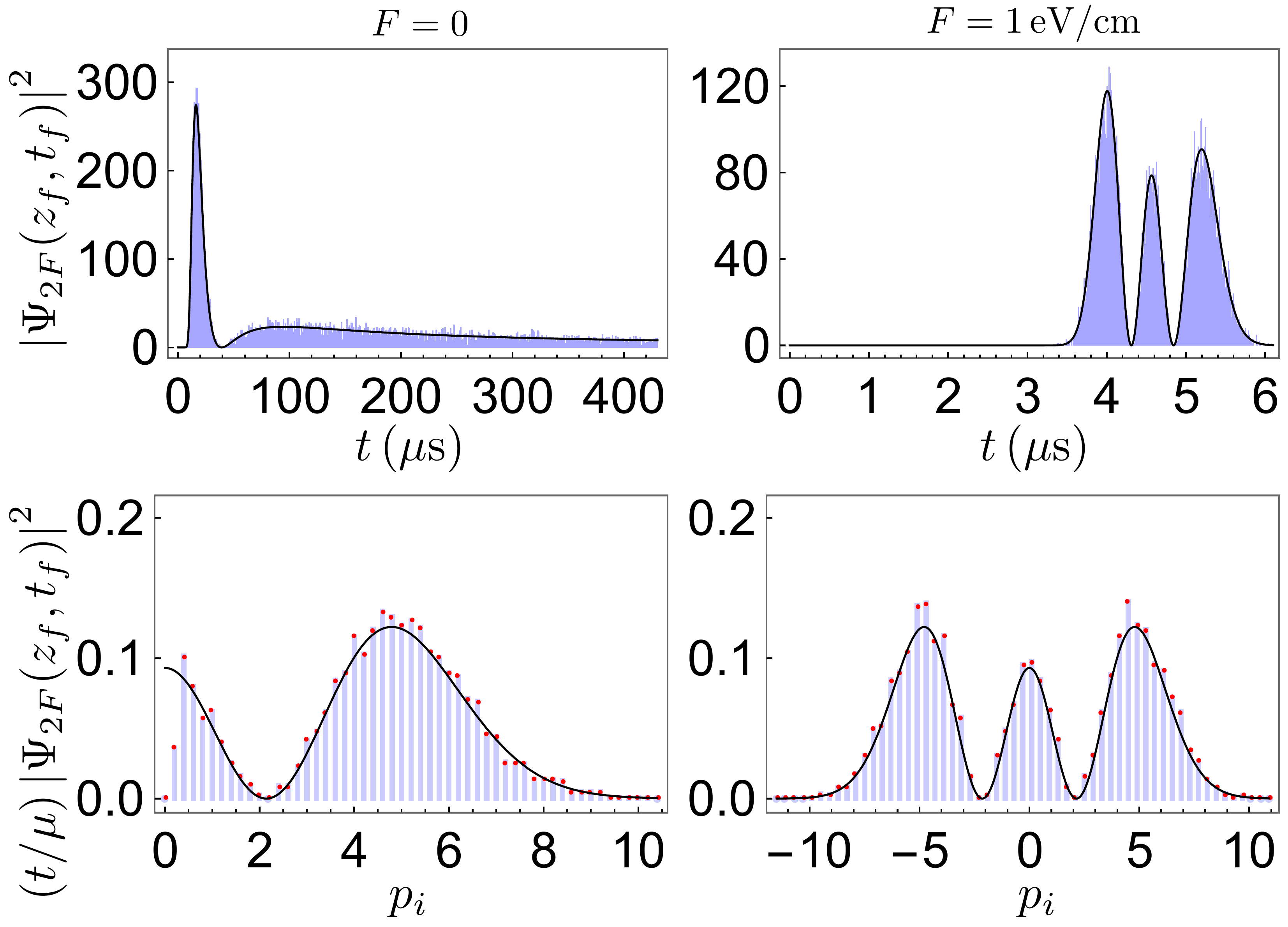}
\caption{\label{fig4} Top row of panels shows histograms of simulated data of the $n=2$ spatial distribution $|\Psi_{2F}(z_f, t_f)|^2$ from Eq.\ (\ref{psi_nt}) using Eq.\ (\ref{PsiF}) time sampled for fixed $z_f = 20 \, \mbox{cm}$. Left panels $F=0$, right panels $F=1 \, \mbox{eV/cm}$. Bottom row of panels shows initial momentum distributions extracted from fits of the histograms using Eq.\ (\ref{phiITsq}) with Eq.\ (\ref{tF}).
In all four panels, the solid black curves show the exact results.}
\end{figure}

Fig.\ \ref{fig4} shows histograms of the simulated time spectra for both $F=0$ and $F = 1 \, \mbox{eV/cm}$. The time axes are given in microseconds ($\mu$s) with bin widths $1 \, \mu\mbox{s}$ in the case of $F=0$ and $0.01 \, \mu\mbox{s}$ for $F\ne0$, readily achievable experimentally.
The electric field acceleration shortens of course the overall time of flight. With electric-field extraction on one sees that the time spectrum resembles more the actual $n=2$ spatial distribution, exactly as we saw in the previous section with a CM boost. In this statistical sense we see that \emph{the wave function fully survives propagation to macroscopic distances.}

Fig.\ \ref{fig4} also shows the initial momentum distributions extracted with piecewise linear fits of the simulated time spectra using Eqs.\ (\ref{phiITsq}) and (\ref{tF}). With the electric field on the full momentum spectrum is recovered.

Finally, one can show using Eqs.\ (\ref{psi_nt}) and (\ref{PsiF}) and $v_f = z_f/t + F t/(2\mu)$ that
\begin{equation}
\frac{j_{nF}(z_f,t_f)}{|\Psi_{nF}(z_f,t_f)|^2 \,  v_f} = 1 + \frac{p_i/p_f }{1+\omega^2 t^2}
\label{equ:jndensity}
\end{equation} 
($p_f \equiv \mu v_f$) independently of the initial state $n$ as before. And as required by the IT, this ratio approaches unity for $\omega t \gg 1$.

\section{Conclusions}
As an alternative to the use of standard scattering theory we have presented the IT in a way which emphasises the possibility of a direct comparison of calculated T-matrix elements with the time spectra measured using modern multi-hit detectors. We have shown how the various forms of the classical action are used to construct the semiclassical wave function in the region immediately outside the reaction zone. Although the IT is an approximation depending upon a stationary phase argument, we have demonstrated a rapid convergence of exact results to the IT form. Indeed, quantum particles propagating from microscopic separations acquire the IT hallmark of classical motion in times and distances which are still of an atomic dimension. Using fragmentation of moving beams or extraction by electric fields we have shown that it is possible to image the full quantum probability distribution of the 
initial momentum state including details of the nodal structure. 

The main result of the IT is to show that the initial quantum probability distribution at the edge of the reaction zone can be viewed as a corresponding classical ensemble of particles which propagate along classical trajectories.  
The familiar spreading of the spatial wave function with time is associated very directly with the ``fanning out" of the ensemble of classical particles of differing momenta emanating from a microscopic volume (cf.\ Fig.\ \ref{fig1p}).
The probability density in coordinate space is related to the probability density in momentum space by the purely classical trajectory-density factor. In our opinion this aspect of the IT represents a far more convincing demonstration of the transition from quantum to classical mechanics than does the Ehrenfest theorem. Furthermore the ensemble rather than the single-particle interpretation of the quantum wave function is at the very basis of the IT picture.
 
In the IT, although the position variables of the wave function describe classical trajectories, the wave function itself remains wholly quantum. All features associated with a quantum wave, e.g.\ interference patterns \cite{BriggsFeagin_ITIII}, are preserved.  In particular, nodal structure is preserved even in external laboratory fields of normal extraction intensity. 

The IT answers in a simple way questions posed as to the visibility or otherwise of nodal structure in measured time spectra. Schmidt et.\ al.\ \cite{Schmidt}, in a notable example of the  capabilities of modern detection techniques, imaged spatially the $H_2^+$ vibrational wave function at the quantum limit. 
 The $H_2^+$ ions were neutralised by electron capture, when the resulting $H_2$ molecules dissociate on a monotonic repulsive potential. Then the initial spatial wavefunction was inferred from the momentum distribution by relating the energy of dissociation to the position on the repulsive curve (the reflection approximation). 
 In the introduction to this paper they comment, 

``However, vibrationally excited molecules have wave functions
with complex structure. In particular, they have nodes
in real space, i.e., positions at which the probability to find
a nucleus is zero even though the molecule vibrates across
these nodes. This fact is rather puzzling to our imagination
guided by classical physics intuition where a particle cannot
move from one to another position without passing all
points in between." 

The IT explains the puzzle by showing that the nodes are visible simply because 
they correspond to a classical ensemble with zero particles occupying the classical trajectory having an initial momentum value at the node. The quantum probability is zero along  a classical path beginning with that particular value of initial momentum. A given particle does not have to traverse nodes as implied in the above quote. The authors also pose the question,

``The question
arises of what the reality of the spatial structure of
vibrational wave functions actually is and whether it can be
actually observed in an experiment given the limits imposed
by the uncertainty principle."

We see that rather than the spatial structure of wave functions  it is the momentum structure that is imaged directly. Nevertheless, nodes are clearly evident. However, there are no limits imposed by the uncertainty principle (although it does play a role in the state preparation) on the propagation described by the IT. From the IT for free propagation, one sees that asymptotically $ \Delta p_f \, \Delta r_f = \Delta p_i \, \Delta r_f \approx  (\Delta p_i)^2 t/\mu$, where $t$ is the time of flight to the detector. 
Hence, although $\Delta p_i$ is of atomic dimensions, $t$ is macroscopic and the uncertainty principle is satisfied by a huge multiple of $\hbar$. This also explains the further dichotomy of the IT as to how classically deterministic position and momentum values can be associated with the existence of a quantum wave function.

\section*{Acknowledgements}
\noindent We much appreciate useful and ongoing discussions with Leigh Hargreaves and Morty Khakoo on all aspects of modern detection technology.
JF acknowledges the ongoing support of the Department of Energy, Chemical Sciences, Geosciences and Biosciences Division of the Office of Basic Energy Sciences.


\begin{thebibliography}{}
\bibitem{energyspectra}
More traditional spectrometers measure energy spectra directly, i.e.\ hits as a function of detected fragment energy. But these can be directly related to time spectra via classical kinematic relations.
\end{thebibliography}
\end{document}